# ADIPOCHONDROCYTES IN RABBIT AURICULAR CARTILAGE

YASSER A. AHMED[1] and MOHAMMED ABDELSABOUR-KHALAF[2]
[1]Department of Histology, Faculty of Veterinary Medicine, South Valley University, Qena, Egypt
[2]Department of Anatomy and Embryology, Faculty of Veterinary Medicine, South Valley University, Qena, Egypt

**ABSTRACT**

Chondrocytes are described as one cell population in different cartilage types. The auricular cartilage in mouse and rat contains unique chondrocytes similar in morphology to white adipocytes and known as "lipochondrocytes". Lipochondrocytes were not mentioned in other species. The current study aimed to explore the existence of this cell type in rabbits. The auricles of adult male white rabbits were harvested and processed for histological examination with light and electron microscopy. With the light microscopy, the auricular cartilage of adult rabbits contained central large rounded adipocyte-like chondrocytes, termed in the current study "adipochondrocytes" The adipochondrocytes were embedded in relatively wide lacunae and had large lipid droplets with a rim of cytoplasm. This result was confirmed by the scanning electron microscopy. With the transmission electron microscopy, the adipochondrocytes showed dark nucleus and electron-dense cytoplasm with few organelles and cytoplasmic processes. The adipochondrocytes of the auricular cartilage in adult rabbits were unique cell type and different from chondrocytes in other cartilage subtypes. This result should be considered during cartilage transplant. Further studies are suggested to investigate the development and physiological roles of adipochondrocytes in the auricular cartilage in rabbits.

## INTRODUCTION

Cartilage is an avascular type of connective tissue exists in three subtypes; hyaline, elastic and fibrocartilage. The cartilage subtypes have different anatomical locations. For example, hyaline cartilage is found on articular surfaces, in the respiratory tract, and in ribs. Elastic cartilage is present in the external ear and epiglottis, while the fibrocartilage predominates in the intervertebral discs, and in the ligaments and tendons (Mescher 2015). Each cartilage subtype was made by a single cell type (chondrocytes) which embedded inside lacunae, and extracellular matrix (ECM), which produced and maintained by the chondrocytes themselves (Ahmed 2007). The ECM consists of glycosaminoglycans (GAGs), proteoglycans and fibers. Based on the predominance of each structure of the ECM, the cartilage three subtypes could be differentiated (Naumann *et al.*, 2002). In general, the elastic cartilage is mostly cellular with a little ECM (Ito *et al.*, 2001). Furthermore, the elastic cartilage in adult rats and mice is unique. Its chondrocytes are occupied by large fat droplets resembling the white adipocytes and its ECM is sparse but rich in elastic fibers and never been mineralized (Sanzone and Reith 1976, Kostovic-Knezevic *et al.*, 1981, Mallinger and Böck 1985, Bradamante *et al.*, 1991). The chondrocytes in the ear pinnae of mice (auricular cartilage) were termed as "Lipochondrocytes" (Sanzone and Reith 1976). Lipochondrocytes were not mentioned in other species. It is not clear if this type of chondrocytes is specific to mice and rats or could be found in the auricular cartilage of other species. We hypothesized that these cells could be found in small mammalian species thus, the current study was undertaken with the aim of exploring these cells in the auricular cartilage of rabbit.

## MATERIALS AND METHODS

**Study area**
The current study was conducted in the Laboratory of Histology Department in the Faculty of Veterinary Medicine, South Valley University, Qena, Egypt from January 2017 to July 2017.

**Sampling**
Apparently healthy, mature (6-month old) male New Zealand White rabbits (n=15) were obtained from the farm of the Faculty of Agriculture, South Valley University, Qena, Egypt. The animals were euthanized with ether then small pieces (0.5 mm) from the auricles were dissected. Some specimens were immediately fixed in either 10% neutral

---

*Corresponding author:* Dr.YASSER A. AHMED
*E-mail address:* yasser.ali@vet.svu.edu.eg
*Present address:* Department of Histology, Faculty of Veterinary Medicine, South Valley University, Qena, Egypt

buffered formalin (NPF; pH=7.4) for 24 hours for the light microscopy or in 2.5% glutaraldehyde for 48 hours at 4C°, then in 1% osmium tetroxide for 2 hours for examination with the scanning (SEM) and transmission (TEM) electron microscopy.

### Light microscopy
NBF-fixed samples were dehydrated in ascending grades of ethanol, embedded in paraffin. Paraffin sections (5μm thickness) were cut with a rotary microtome (Leica RM2255) and stained with hematoxylin and eosin (H&E) as a general stain, periodic acid–schiff (PAS) for neutral GAGs, alcian blue for acidic GAGs, combined PAS-alcian blue for both types of GAGs, safranin-O for proteoglycans, Masson's trichrome for collagen fibers and orcein stains for elastic fibers detection (Jones *et al.* 2008). Sections were examined with the light microscope (Leica DMLS). Photographs were captured with Leica digital camera (Leica ICC50) using ×4, ×10, ×40 and ×100 objectives in JPEG format.

### Scanning Electron microscopy
Fixed specimens were dehydrated in ascending grades of ethanol. The specimens were dried, coated with gold in a sputter coater and examined with the SEM (JEOL 5500) at the Central Laboratory of the SVU.

### Transmission Electron microscopy
Fixed specimens were dehydrated in ascending grades of ethanol and embedded in Spurr's resin. Semithin and ultrathin sections were taken with a glass knife. Semithin sections (0.5 μm thickness) were stained with toluidine blue and ultrathin sections (80-100 nm) were stained with lead citrate and uranyl acetate, then examined with the TEM (JEOL1010) at the Central Laboratory of the SVU.

### Histomorphometry
The thickness of the auricular cartilage, the diameters of the lacunae and lipid droplets of adipochondrocytes, the adipochondrocyte/ ECM ratio area and cellular density (the number of cells in a specific area) were measured using the free imageJ software (https://imagej.nih.gov/ij/). The results were analyzed by EXCELL 2016 and expressed as mean ±SE.

## RESULTS

### Morphology of the adipochondrocytes
The auricular cartilage plate of the white rabbits consisted of hypertrophic white adipocyte-like chondrocytes inside their rounded to oval lacunae. These cells were termed "Adipochondrocytes" in the current study. The adipochondrocytes occupied most of the central zone of the auricular cartilage and were separated from each other by a sparse ECM and surrounded by flattened collagenous tissue of perichondrium. In addition, smaller ovoid cells were squeezed between the central hypertrophied adipochondrocytes and the perichondrium, and contained varying amounts of lipid droplets (Fig. 1A-C). The cytoplasm of adipochondrocytes was mostly occupied by single large rounded to oval lipid droplets. The lipid droplets of the adipochondrocytes dissolved during paraffin section preparation, leaving empty spaces surrounded by a rim of cytoplasm, resembling the "signet ring" appearance of the adipocytes (Fig. 1A, B). With osmium tetroxide-fixed specimens, the lipid could be preserved and appeared as large droplets nearly filling the cell except for a rim of toluidine blue-stained cytoplasm (Fig. 1C). The adipochondrocytes had flattened peripherally located nucleus and some cells were binucleated (Fig. 1B). The perichondrial or territorial matrix appeared empty in sections stained with H&E (Fig. 1A, B) or contained some metachromatic materials with toluidine blue (Fig. 1C).

With the SEM, the overall structure of the cartilage revealed its appearance seen with the light microscopy. Adipochondrocytes contained large lipid droplets. Some adipochondrocytes had many very small lipid droplets scattered over the cytoplasm and some had large and small droplets, while some cells appeared empty. Furthermore, some lacunae appeared closed or partially opened (Fig. 1D, E).

The TEM examination revealed lipid droplets surrounded by amorphous dark areas in the electron-dense cytoplasm and dark nucleus. Cytoplasm contained sparse organelles except for RER, Golgi apparatus, and secretory granules, and the plasma membrane showed many cellular processes extended into a proteoglycan-rich territorial matrix (Fig. 1F).

### Morphology of the cartilage ECM
A territorial ring of matrix surrounded the adipochondrocytes appeared empty with H&E (Fig. 1A, B), stained metachromatic with toluidine blue (Fig. 1C) and stained strongly positive with alcian blue and positive with PAS (Fig. 2A) and safranin-O (Fig. 2B).

With TEM, the territorial matrix was well preserved and characterized by its high contents of proteoglycan materials (Fig. 1F). The interterritorial matrix between the adipochondrocytes stained positive with PAS (Fig. 2A), safranin-O (Fig. 2B), Masson's trichrome (Fig. 2C) and orcein stains (Fig. 2 D) but weak positive with alcian blue (Fig. 2A). With the TEM, the interterritorial matrix appeared consisted of a network of dense-granule proteoglycans integrated with amorphous areas of elastin and collagen fibers were seen in some areas (Fig. 2 E, F).

### Histomorphometry
The thickness of the auricular cartilage of adult rabbits was 118.1 ± 1.8μm. The adipochondrocytes embedded in lacunae with a diameter of 32.9 ± 0.8μm and average diameter ranged between 17μm (minimum) and 54.5μm (maximum). The adipochondrocytes contained lipid droplets with a diameter of 20.5 ± 0.5μm and ranged between 12.5μm (minimum) and 28.6μm (maximum). The adipochondrocyte/ECM ratio was about 49.2% and cellular density was about one cell/104.9 μm.

**Figure legends**

**Figure 1: Morphology of adipochondrocytes in adult rabbit auricular cartilage**

A-C: Paraffin sections stained with H&E (A, B), and a semithin section stained with toluidine blue (C). D-F: SEM (D, E) and TEM (F) micrographs of rabbit auricular cartilage. A: Outline of auricular cartilage; perichondrium (p) and lipid droplet of adipochondrocytes (L). B: Adipochondrocytes; lipid droplets (L), perichondrial matrix (pm), ovoid cells (arrowhead), perichondrium (p) and interterritorial ECM (m). C: Osmium tetroxide-preserved lipid droplets-preserved lipid droplets (L), metachromatic perichondrial matrix (pm), ovoid cells (arrowhead) under perichondrium (p). D: Auricular cartilage outline with SEM; perichondrium (p) and adipochondrocytes with lipid droplets (L) in the cytoplasm (c); note small lipid droplets (s), varying sizes lipid droplets (v), cells with no lipids (e), opened (o), partially opened (large arrowhead) and closed (small arrowhead) lacunae. E: Adipochondrocyte inside lacunae (Lc); a large lipid droplet (L) filling the cytoplasm (c). F: Adipochondrocyte with a lipid droplet (L), amorphous materials in a dark cytoplasm (c), dark nucleus (N), Golgi apparatus (black arrowhead), secretory granules (white arrowhead), RER (white arrow), cytoplasmic processes (black arrow) and perichondrial matrix (pm). Bars = 100 μm (A), 40 μm (B, C), 50 μm (D), 10 μm (E) and 0.5 μm (F).

**Figure 2: Morphology of the ECM of adult rabbit auricular cartilage**

Light micrographs of paraffin sections (A-D) and electron micrographs (E, F) from adult rabbit auricular cartilage. A-D: Cartilage interterritorial ECM (m) positive with PAS-alcian blue (A), safranin-O (B), Masson's trichrome (C) and orcein (D); note lipid droplets of adipochondrocytes (L). White arrowheads indicate elastin of the matrix (m) in E, F. Note, lipid droplet (L) surrounded by amorphous dark cytoplasm (c) of adipochondrocytes and perichondrial matrix (pm) in E and collagen fibers (black arrowhead) in F. Parts A-D have the same magnifications; bar = 40 μm and parts E and F have the same magnifications; bar = 0.5 μm.

**Figure 1**

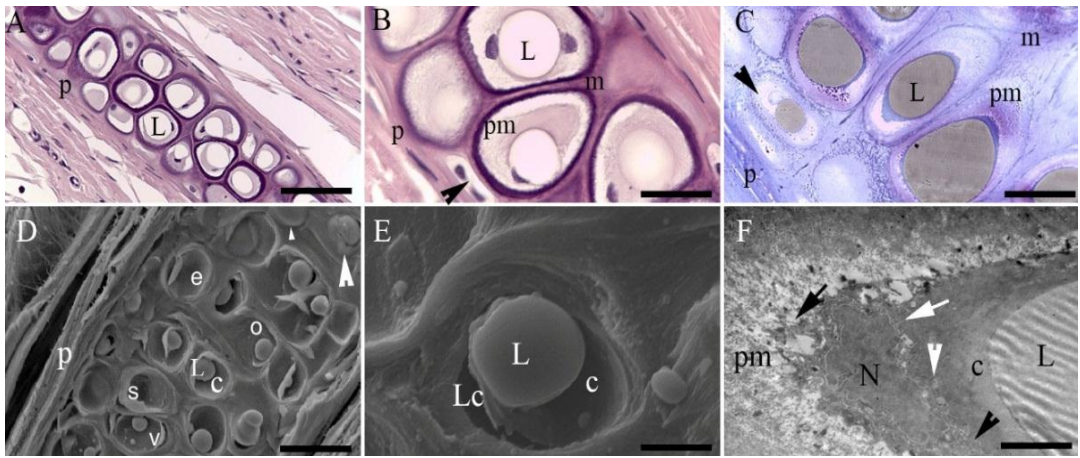

**Figure 2**

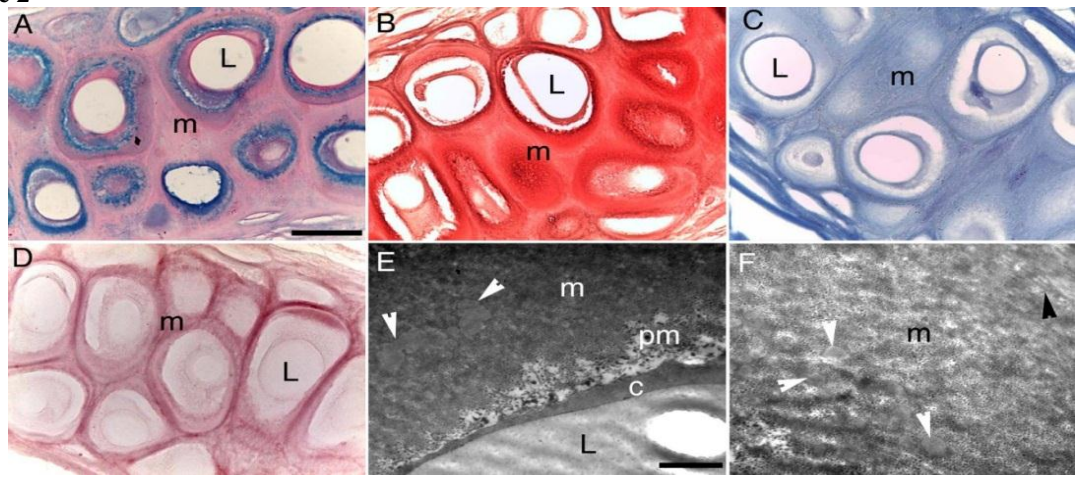

# DISCUSSION

Chondrocytes are responsible for building and maintaining the different subtypes of cartilage. Chondrocytes are commonly described as one cell type by many authors (Karim and Hall 2017, Li *et al.*, 2017). The early studies described that chondrocytes in the auricular cartilage of mice and rats are unique, as their cytoplasm is mostly occupied by large lipid droplets and were termed "lipochondrocytes" (Sanzone and Reith 1976). This type of chondrocyte was not mentioned in the subsequent publications or in other species. We hypothesized the existence of this type of chondrocytes in small animals. The current study was carried out with the aim of rediscovering this type of chondrocytes in rabbits. Auricular cartilage from adult white New Zealand rabbits was harvested and processed for light and electron microscopic examinations.

The study showed that chondrocytes with a morphological appearance of adipocytes were existing in the rabbit auricular cartilage. These cells were termed as "adipochondrocytes" in the current study and not lipochondrocytes as termed in the early studies (Sanzone and Reith 1976). We preferred to use the term "adipochondrocytes" because of cells have a morphological appearance very similar to that of the white adipocytes. However, the term "lipochondrocytes" comes from the high lipid contents of these cells and not from their morphological similarity with the adipocytes. Many cell types contain varying amounts of lipids but don't resemble the white adipocytes such as hepatocytes (Canbay *et al.*, 2007) and Leydig cells (Ahmed *et al.*, 2012) and it was not applicable to call these cells "lipo-hepatocytes" and "lipo-Leydig cells" cells by the same way Sanzone and Reith (1976) used the term "lipochondrocytes". Thus, the term adipochondrocytes was more expressive of the morphology of the chondrocytes seen in the current study. The physiological role of the lipids in adipochondrocytes has not been described yet. Some authors suggested that the lipid granules indicate degenerative changes in these cells (Nielsen 1976). The chondrocytes in cartilage are known to die by non-apoptotic modes of cell death (Roach *et al.*, 2004, Ahmed *et al.*, 2007b). However, no evidence or example of such cellular death could be found in the current study. We expected the lipids act as a source of stored energy as lipids in adipocytes do. This expectation was excluded by (Kostovic-Knezevic *et al.*, 1981), when they mentioned that the lipid contents of "adipochondrocytes" in auricles have not been changed after the loss of 50% of body weight due to starvation. It is likely to suggested that the lipids of adipochondrocytes may be important for the mechanical support of the auricles of the external ear in such species. Adipochondrocytes contained electron-dense cytoplasm with dark peripheral nucleus and sparse organelles represented by RER, Golgi apparatus and many secretory granules, however it looks active in proteoglycan-rich ECM secretion. Furthermore, the amorphous materials around the lipid droplets seen in this study were similar to those seen in rat auricular cartilage and were described as microfilaments (Kostovic-Knezevic *et al.*, 1981). The adipochondrocytes appeared to share some characteristic features of hypertrophic dark chondrocytes described in the articular cartilage from different species (Roach *et al.*, 2004, Ahmed *et al.*, 2007a, Ahmed *et al.*, 2007b, Chen *et al.*, 2010a, Chen *et al.*, 2010b) such as electron-dense cytoplasm and developed RER and Golgi apparatus with many secretory granules and cellular processes. Like, hypertrophic dark chondrocytes in the growth cartilage, the adipochondrocytes in the rabbit's auricular cartilage are active in GAGs and proteoglycan synthesis as confirmed by PAS-alcian blue and safranin-O staining and by TEM examination.

In conclusion, the adipochondrocyte, a cell type share some morphological features of adipocytes and dark hypertrophic chondrocytes, could be identified in the auricular cartilage of adult rabbits and appeared active in the ECM synthesis. Further studies are required for further exploring molecular and functional features of adipochondrocytes. It is known that chondrocytes from different locations behave differently in tissue-engineered cartilage models. Thus, this study is not only important for a better understanding the morphology of different cartilage types, but also could be important for improving the cartilage tissue engineering models.

# ACKNOWLEDGEMENTS

The authors would like to thank Prof. Antar Abdallah, Professor of English education at Taibah University, Saudi Arabia, for English proofreading of the manuscript.

## الخلايا الدهنوغضروفية "Adipochondrocytes" فى غضروف صوان الاذن فى الارانب

### ياسر عبد الجليل أحمد ، محمد عبد الصبور احمد خلف

E-mail: yasser.ali@vet.svu.edu.eg    Assiut University web-site: www.aun.edu.eg

تعتبر الخلايا الغضروفية هي الخلايا المكونة للأنواع المختلفة من الغضاريف، حيث أن غضروف صوان الأذن في الفئران والجرذان يتكون من خلايا غضروفية تشبه إلى حد كبير الخلايا الدهنية لذا أطلق عليها الخلايا الدهنوغضروفية ولكنها لم توصف في الفصائل الأخرى، لذلك تهدف هذه الدراسة إلى البحث عن مثل هذه الخلايا في غضاريف الأذن في الأرانب. تم جمع العينات من صوان الأذن من ذكور الأرانب النيوزلاندية البيضاء البالغة، وتم تجهيز العينات لفحص التركيب الهستولوجى باستخدام الميكروسكوب الضوئي والميكروسكوب الالكتروني. أظهرت النتائج أن الخلايا المكونة لغضاريف الأذن في الأرانب تتميز بأنها خلايا دائرية مركزية كبيرة الحجم تشبه الخلايا الدهنية لذا أطلق عليها في الدراسة الحالية " الخلايا الدهنوغضروفية". وصفت هذه الخلايا بأنها توجد داخل فجوات كبيرة وتحتوى على كمية كبيرة من الدهون مع وجود حافة بسيطة من السيتوبلازم. وقد تم تأكيد النتائج من خلال الميكروسكوب الالكتروني الماسح والنافذ الذي أوضح أنها تحتوى على نواة داكنة وسيتوبلازم داكن مع وجود عضيات قليلة العدد وزوائد سيتوبلازمية. من خلال هذه النتائج يتبين أن الخلايا الغضروفية الدهنية المكونة لغضروف صوان الأذن في الأرانب البالغة تختلف عن الخلايا الغضروفية المكونة للأنواع الأخرى من الغضاريف. هذه الدراسة يجب الاعتبار بها عند زراعة الغضاريف. يوصى بعمل دراسات إضافية لمعرفة كيفية تطور ووظائف الخلايا الدهنوغضروفية في الأرانب.